\documentclass[namedreferences]{solarphysics}
\usepackage[optionalrh,natbib]{spr-sola-addons} % For Solar Physics 
\usepackage{graphicx}        % For eps figures, newer & more powerfull
\usepackage{amssymb}        % useful mathematical symbols
\usepackage{color}           % For color text: \color command
\usepackage[T1]{fontenc}
\usepackage{url}             % For breaking URLs easily trough lines
\usepackage{breakurl}
\usepackage[psamsfonts]{amsfonts}
\usepackage{mathptmx}
\usepackage{booktabs}
\definecolor{darkgreen}{RGB}{0,142,128}

\newcommand{\changed}[1]{{#1}}
\usepackage[pdfborder={0 0 0},urlcolor=blue,breaklinks,colorlinks,linkcolor=blue,citecolor=blue]{hyperref}

\newcommand{\etal}{{\it et al.}}
\renewcommand{\vec}[1]{{\mathbfit #1}}

\newcommand{\aap}{    {\it Astron. Astrophys.}}

\newcommand{\apj}{    {\it Astrophys. J.}}
\newcommand{\apjl}{   {\it Astrophys. J. Lett.}}

\newcommand{\epjb}{   {\it European Phys. J. }}

\newcommand{\grl}{    {\it Geophys. Res. Lett.}}

\newcommand{\jgr}{    {\it J. Geophys. Res.}}

\newcommand{\pra}{    {\it Phys. Rev. A}}
\newcommand{\pre}{    {\it Phys. Rev. E}}
\newcommand{\prl}{    {\it Phys. Rev. Lett.}}
\newcommand{\solphys}{{\it Solar Phys.}}

\graphicspath{{Figs/}}

\begin{document}

\begin{article}

\begin{opening}

\title{Deterministically Driven Avalanche Models of Solar Flares}
\author{Antoine Strugarek \sep Paul Charbonneau \sep Richard Joseph \sep
  Dorian Pirot$^{1}$}
\institute{D\'epartement de physique, Universit\'e de Montr\'eal, C.P. 6128 Succ. Centre-Ville,
Montr\'eal, Qc, H3C-3J7, CANADA \\ email: strugarek@astro.umontreal.ca
\\$^1$ also at Universit\'e de Rennes {I}, Rennes, France}
\date{\today}

\runningauthor{A. Strugarek \textit{et al.}}
\runningtitle{Deterministically Driven Avalanche Models of Solar Flares}

%\maketitle

\begin{abstract}
  We develop and discuss the properties of a new class
  of lattice-based avalanche models of solar flares. 
  These models are readily amenable to a relatively unambiguous
  physical interpretation in terms of slow twisting of a coronal loop.
  They share similarities with other avalanche models, such
  as the classical stick--slip self-organized
  critical model of earthquakes, in that they are driven globally by
  a fully deterministic energy loading process. The model design leads
  to a systematic deficit of small scale avalanches.
  In some portions of model space, mid-size and large avalanching
  behavior is scale-free, 
  being characterized by event size distributions that have the
  form of power-laws with index values, which, in some parameter
  regimes, compare favorably to 
  those inferred from solar EUV and X-ray flare data. For models using
  conservative or near-conservative
  redistribution rules, a population of large, quasiperiodic
  avalanches can also appear. Although without direct counterparts
  in the observational
  global statistics of flare energy release, this latter behavior
  may be relevant to recurrent flaring in individual coronal
  loops. \changed{This class of models could provide a basis for the prediction
  of large solar flares}.
\end{abstract}


\end{opening}

\section{Reconnection Avalanches and Parker's Nanoflare Hypothesis}

Identifying the physical mechanism responsible for
coronal heating is still standing as one of the grand
challenges of contemporary solar physics \citep[see][for different
theoretical and observational
viewpoints]{Klimchuk:2006ka,Aschwanden:2007kq,DePontieu:2011fk}.
Nearly a quarter
of a century ago, \citet{Parker:1983hf} began exploring in earnest
the possibility that coronal heating could be due to 
thermal energy released by small magnetic reconnection events occurring
continuously throughout the magnetized corona
\citep[see][and references therein for related physical scenarii]{Parker:1983gr,Parker:1988ia,Priest:2002hx}.
His physical picture for nanoflare production is now
well-known. Photospheric fluid motions associated
with turbulent convection randomly shuffle the magnetic
footpoints of coronal loop leading to a braiding
of fieldlines throughout the volume of the loop, with which is
associated an inexorable buildup of current sheets, as
individual fieldlines try to return to dynamical equilibrium
under the topological constraints imposed by flux-freezing. As these
localized current systems build up, plasma instabilities eventually
set in and lead to energy release
through magnetic reconnection. Parker dubbed these 
elementary energy release events ``nanoflares'',
went on to argue that they occur continuously throughout 
the corona, and conjectured that they are responsible
for coronal heating.

%The energy flow is quite noteworthy in this scenario. The 
%thermal energy liberated throughout the corona by the weak but
%numerous and ever-present nanoflares ultimately originates from
%the kinetic energy of photospheric convective flows. These flows
%do work against the magnetic field as footpoints are randomly
%forced, and the associated energy is carried up into the corona as Alfv\'en waves
%travelling along the length of the loop, stored temporarily
%in electric current systems associated with the unavoidable
%tangential discontinuities forming where fieldlines kink about
%one another, to be finally released locally as thermal
%energy in the course of the reconnection process.

Intense efforts have been expended to observationally verify or contradict
Parker's conjecture. This is no easy task, as it involves measuring the
energy output of flares
very close to the detection limit of the best current solar imagers
in space. On the other hand, decades of flare observations have 
shown that the probability distribution function [$f(E)$]
for flare energy [$E$] takes the form of a
power law [$f(E)\propto E^{-\alpha}$] spanning
some eight orders of magnitude in energy, with
$\alpha$ estimated in the range $1.65$ -- $1.95$
\citep[\textit{e.g.}][and references
therein]{Aschwanden:2000gc,Aschwanden:2002fl,Aschwanden:2011hw}.
This is too low for small flares to dominate energy input
into the corona (requiring $\alpha>2$), unless the PDF steepens abruptly below
the current detection limit. This is possible but increasingly unlikely
as instruments such as the EUV imagers on the \textit{Transition
Region And Coronal Explorer} (TRACE) and \textit{Solar Dynamics
Observatory} (SDO) have pushed 
the detection limit right up to the boundary of the nanoflare
regime. Another possibility is that physical models used to infer thermal energy release
from EUV or soft X-ray fluxes are in error, due to erroneous
assumptions regarding the intrinsic geometry of the flaring volume
\citep{McIntosh:2001jt}, or of the physical
conditions therein, required
to infer thermal energy release from observed radiative fluxes
\citep[see][and references therein]{McIntosh:2000kn,Aschwanden:2002jy,Aschwanden:2013kb}.

Whatever recent observational determination of $f(E)$
may imply for coronal heating,
Parker's scenario
certainly remains a fine model for solar flares in general.
Although not originally emphasized by Parker, his nanoflare
scenario contains all ingredients now understood to be required
to lead to self-organized criticality \citep[hereafter SOC; see,
\textit{e.g.}][]{Bak:1987ef,Jensen:1998ww}: an open dissipative
system -- here a coronal loop -- slowly driven
-- here by random footpoint motions -- and subjected to a self-limiting
local threshold instability -- here magnetic reconnection.
Despite the slow and gradual loading of energy,
such a physical system will release energy in an
intermittent manner, in the form of
avalanches that can span up to the whole system.

\citet{Lu:1991hp} have proposed
an avalanche-type model for solar flares that
has become a reference model in the solar context, although
numerous variations have now been developed over the intervening years.
The present article describes one such variation that (we believe) is
particularly interesting in that i) it is readily amenable to
(relatively) unambiguous physical
interpretation in terms of twisting of a coronal loop, and ii) it can
yields qualitatively distinct flaring behavior as the degree of conservation %driving amplitude
is allowed to vary.
Section \ref{sec:newmod} briefly describes the original Lu and Hamilton
model \citep[more specifically, the version put forth by][hereafter
the ``LH model'']{Lu:1993cy},
along with the various manners in which we have altered the forcing
and redistribution rules. Section \ref{sec:results} provides
modelling results, comparing and contrasting the statistical
properties of avalanches in the LH model and our
variations thereof. We conclude in Section \ref{sec:conclusion} with
a discussion comparing and contrasting our models with other
classes of globally and deterministically driven avalanches
models developed in different physical contexts, outlining
some possible model improvements,
and by summarizing the implications of our modelling results
for coronal heating (and more specifically with regards
to the flaring behavior of small coronal loops) as well as solar flares predictions.

\section{A Class of Avalanche Models With Deterministic Driving}
\label{sec:newmod}

\subsection{The Lu and Hamilton Model \label{ssec:LH}}

Following in part \citet{Kadanoff:1989ki}, \citet{Lu:1991hp}
developed
a SOC avalanche model for solar flares that by now has become 
a kind of ``standard'' \citep[for a review, see][]{Charbonneau:2001fv}.
The original Lu and Hamilton
model is a cellular automaton defined over a
3D regular cartesian grid with nearest neighbour connectivity
(top+down+right+left+front+back) over which a vector field
[$\vec{A}$] is defined. Here we will consider instead a 2D version
of the model, where a
scalar nodal quantity [$A_{i,j}^n$] is defined
over the lattice. \citet{Robinson:1994cf} has shown that for the type
of driving used in the Lu and Hamilton  model, the use of a vector nodal
variable yields results identical (statistically)
to the use of a scalar variable. Also,
strictly speaking the term ``cellular automaton'' should be restricted
to lattice models where the nodal variable is discrete (\textit{i.e.} integers
mapping to a finite
number of possible nodal states); in cases where the nodal variable is
continuous, one is dealing with a ``coupled map lattice''. We retain
``cellular automaton'' here, because it has become common usage
in this context.
The superscript $n$ is a discrete time index, and the subscript pair
[$i,j$] identifies a single node on the 2D lattice.
Keeping $A=0$ on the lattice boundaries, the cellular automaton is
driven by adding one small increment in $A$ per time step,
at some randomly selected node that changes from one time step to the
next.  A deterministic stability criterion is defined
in terms of the local curvature of the field at node $(i,j)$:
\begin{equation}
  \label{eq:stab1}
  \Delta A_{i,j}^n\equiv  A^n_{i,j}
  -\frac{1}{4}\sum_kA_k^n \, ,
\end{equation}
where the sum runs over the four nearest neighbours at
nodes $(i,j\pm 1)$ and $(i\pm 1,j)$.
If this quantity exceeds some pre-set threshold [$Z_c$]
then an amount of nodal variable [$Z$]
is redistributed to the same four nearest neighbours
according to the following discrete, deterministic rules:
\begin{equation}
  \label{eq:redis1}
  A_{i,j}^{n+1}= A_{i,j}^n-\frac{4}{5} Z~,
\end{equation}
\begin{equation}
  \label{eq:redis2}
  A_{i\pm 1,j\pm 1}^{n+1}= A_{i\pm 1,j\pm 1}^n+\frac{1}{5} Z~,
\end{equation}
with $Z=\rm{sign}(\Delta A)Z_c$.
Following this redistribution, it is possible that one of the nearest neighbour nodes
now exceeds the stability threshold.
The redistribution process begins anew from this node, and so on in
classical avalanching manner. Driving is suspended during avalanching,
implicitly implying a separation of timescales between driving and
avalanching dynamics, and all nodal values are updated synchronously
during an avalanche to avoid introducing a directional bias in
avalanche propagation.

It is readily shown that these
redistribution rules, while conservative in $A$, lead to a decrease
in $A^2$ summed over the five nodes involved by an amount
\begin{equation}
  \label{eq:erel1}
  \Delta e_{i,j}^{n}=\frac{4}{5}\left(2{\left|\Delta A_{i,j}^n\right|\over Z_{c}}-1\right)Z_{c}^2 ~,
\end{equation}
with the energy released being ``assigned'' to the unstable node $(i,j)$.
If one identifies $A^2$ with a measure of magnetic energy (more
on this shortly), the total
energy liberated by all unstable node at a given iteration
is then equated to
the energy release per unit time in the flare:
\begin{equation}
  \label{eq:elatt}
  (\Delta E)^n=\sum_{\rm unstable}\Delta e_{i,j}^{n}~. 
\end{equation}
A natural energy unit
here, used in all that follows, is the quantity of energy [$e_0$]
liberated by a single node exceeding
the stability threshold by an infinitesimal amount;
setting $|\Delta A_{i,j}|\approx Z_{c}$ in Equation~(\ref{eq:erel1})
immediately leads to
\begin{equation}
  \label{eq:eunit}
   e_0=(4/5)Z_c^2~.
\end{equation}
This very simple model yields a strikingly good representation
of flare statistics, namely the observed power law form
(and associated exponents) of the frequency distributions
of flare peak energy release [$P$], duration [$T$], and total energy release
\citep[see][]{Lu:1991hp,Charbonneau:2001fv,Aschwanden:2002fl}. However,
Equations~(\ref{eq:stab1}) -- (\ref{eq:redis2})
are admittedly very far removed
from the MHD equations governing the process of magnetic reconnection,
but there exist plausible bridges across this conceptual chasm.

\subsection{Physical Interpretation\label{ssec:phys}}

We adopt here a variation \citep[][]{Charbonneau:2013vt} of the
interpretative picture 
originally put forth by LH. %  and further
% elaborated by others (\textit{e.g.} Isliker \etal~2000, 2001).
The 2D lattice will be considered as a slice
across a coronal loop, \textit{i.e.} in cylindrical coordinates
$(\varpi,\phi,z)$ (see schematic in Figure \ref{fig:schematic_global_forcing}A),
a slice in the $\varpi\phi$-plane, with
the loop axis in the $z$-direction, and the lattice variable
as the $z$-component of the magnetic vector potential
$\vec{A}$ such that $\vec{B}=\nabla\times \vec{A}$.
Such a magnetic field automatically satisfies the solenoidal
constraint $\nabla\cdot \vec{B}=0$.
The vector potential component [$A_z$] then defines
the magnetic field component in
the $\varpi\phi$-plane, which can be directly related to the degree
of twist in the loop, \textit{i.e.} the ratio of transversal to
longitudinal field components. As argued by LH,
adding small random increments of $A$ at one lattice node
then corresponds to adding a small
amount of twist somewhere in the loop, which fits nicely
the Parker scenario outlined above. Moreover, the stability
criterion becomes a threshold condition on $\nabla^2(A_z \hat {\vec{z}})$,
\textit{i.e.} the magnitude of the longitudinal component of the
electric current, which is also satisfying from the point of
view of MHD stability.

This {\it ansatz} suffers from one significant difficulty,
however: there is not necessarily a one-to-one correspondence between
the squared nodal variable \changed{(squared vector potential)} and magnetic energy as conventionally
defined [$\propto \int B^2{\rm d}V$]. It was nevertheless shown \citep[see Figure 12.6 in][]{Charbonneau:2013vt}
that the correspondance of $A^{2}$ to
$B^{2}$ is empirically almost always satisfied\changed{, in the sense that
the variations of $A^{2}$ almost always follow the variations of $B^{2}$ in
our model}.

\subsection{Deterministic Global Driving\label{ssec:detdrive}}

\begin{figure}
  \centering
  \includegraphics[width=\linewidth]{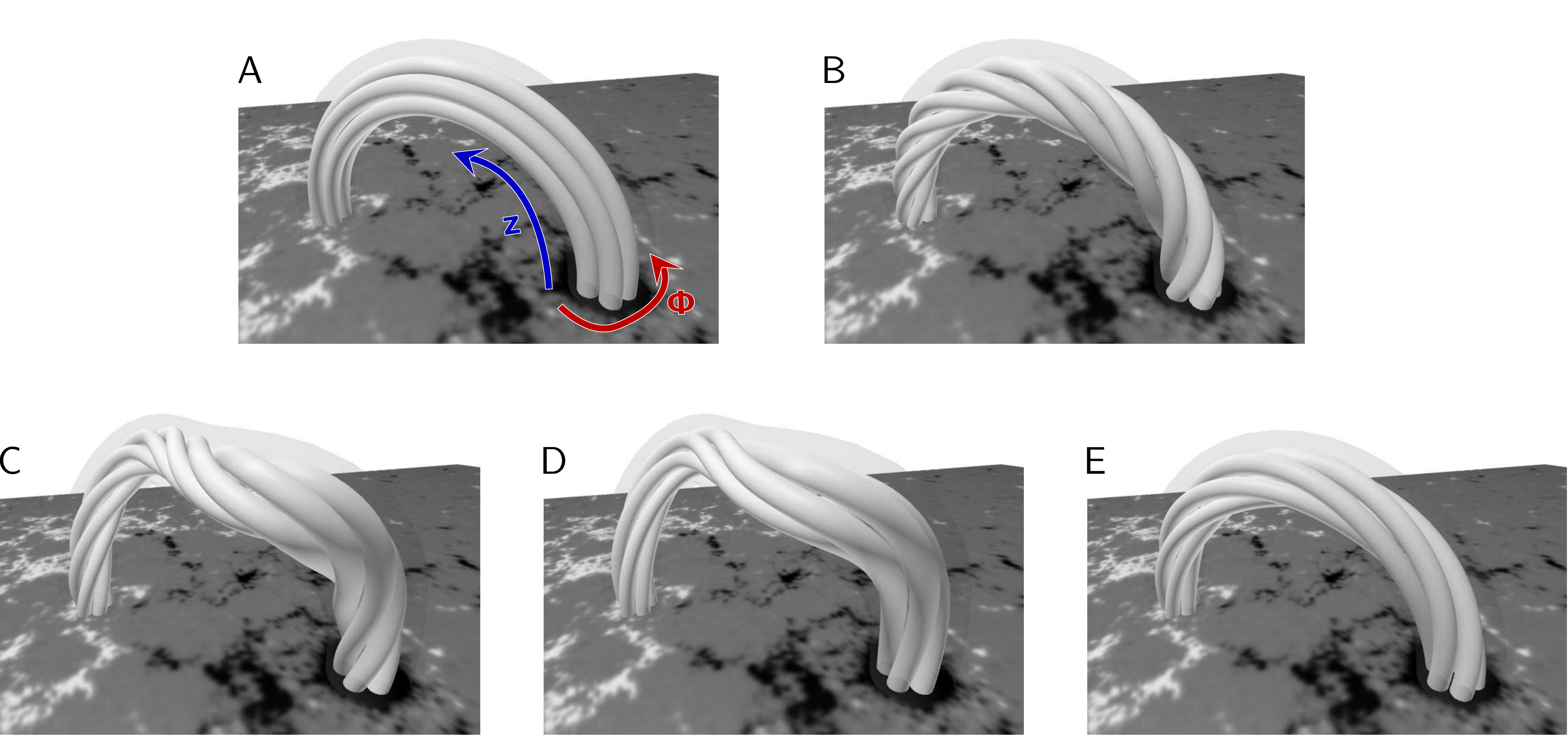}
  \caption{Cartoon of a physical interpretation of the deterministic
    forcing and its associated avalanches in the context of a solar
    coronal loop. See text for a detailed description.}
  \label{fig:schematic_global_forcing}
\end{figure}

The Parker picture of random shuffling of a loop's magnetic
footpoint by photospheric flows makes sense if the loop
is of diameter comparable to (or preferably larger) than the granular
length scale, and if no larger fluid motions co-exist near the base of
the loop. % For loop diameters smaller than this scale
The granular or larger scale flow otherwise displaces the footpoints
in a spatially coherent manner far
removed from random shuffling. One particularly interesting form
of such global forcing is a twisting of the loop's footpoints
(illustrated in Figure \ref{fig:schematic_global_forcing}),
which then propagates upwards and accumulates along the length
of the loop [$A\rightarrow B$]. Such a twisted magnetic loop can
become unstable \citep[e.g. to a kink-type instability, $B\rightarrow C$, see][and
references therein]{Baty:1996we,Bareford:2010ka}. Local tangential discontinuities
arise and associated current sheets appear. Magnetic reconnections
release energy locally and reduce the effective stresses [$C\rightarrow D$], 
possibly restoring the loop stability [$D\rightarrow
E$]. \changed{Note also that the final stable state [$E$] is not
  necessarily twist-free; it is generally composed of a particular pattern of
  stresses resulting from the past reconnection history in the loop.}

This form of global forcing and energy release mechanism has a direct equivalent
in our 2D cellular automaton. Working in cylindrical coordinates
$(\varpi,\phi,z)$, with the loop axis in the $z$-direction,
the total magnetic field can be written as
\begin{equation}
\label{eq:bdef}
\vec{B}(\varpi,\phi,t)=\nabla\times (A_z(\varpi,\phi,t){\hat \vec{z}})+B_z(\varpi){\hat \vec{z}}~.
\end{equation}
Identifying now the nodal variable $A$ with $A_z$ sampled on the
lattice, consider the following deterministic, global driving mechanism:
\begin{equation}
  \label{eq:detdrive}
  A_{i,j}^{n+1}= A_{i,j}^n \times (1+\varepsilon)~,
\qquad 0<\varepsilon\ll 1~,\qquad \forall\,(i,j)~,
\end{equation}
where the parameter $\varepsilon$ ($\ll 1$) is a measure of the
driving rate.
With $A_z=0$ on the boundaries, the SOC state is peaked at 
the center of the lattice, and far enough from the latticed boundaries
is characterized by near-axisymmetry about the lattice center.
This implies that $\nabla\times (A_z{\hat{\vec{z}}})$ is primarily in the $\phi$-direction.
The above driving mechanism thus amounts to a geometrical increase
of $B_\phi$ by a factor $(1+\varepsilon)^n$ after $n$ driving iterations,
while $B_z$ remains unaffected. In other words, the degree of
twist within the loop increases geometrically as well. The stability criterion
based on the magnitude of the longitudinal component of the
electric current then becomes a natural criterion on the intensity of
the current sheet(s) developing in the loop.

In addition, $B_\phi$ is increasing in direct proportion to $A_z$
everywhere in the loop, which means that
associating $A^2$ summed over the lattice with magnetic energy is
entirely realistic in this context. All avalanche models discussed
in the remainder of this article make use of this global,
deterministic driving scheme.
As in the LH model, driving is interrupted during avalanching, which
amounts to assuming that the driving timescale is much longer than
the avalanching timescale, a reasonable assumption in the solar coronal context
\citep[for a discussion, see][]{Lu:1995ce}.

Similar global, deterministic driving schemes have been used extensively
in the context of SOC models of seismic faults, based on a cellular
automaton representation of the Burridge--Knopoff stick--slip model
\citep[see][and references therein]{Olami:1992do,Lise:2001du}. The
main difference from our model resides in their stability criterion
that applies to the nodal variable rather than, in our case, to its curvature.
A global, deterministic driver
has also been used by \citet{Liu:2006cp,VallieresNollet:2010ed,Liu:2011jr} in their
cellular automaton model of magnetospheric substorms generation
in the Earth's central plasma sheet. \changed{To our knowledge,
  \citet{Hamon:2002hs} have been the only authors to explore
  deterministically driven avalanche models in the context of solar
  flares. They used a slightly modified version of the original model
  of \citet{Olami:1992do}, in which the threshold triggering the
  avalanching behavior applies is defined on the nodal variable
  rather than its curvature. In the present study, we use the latter
  criterion, which is naturally inherited from the original LH model.}
% However, most variations on the LH models studied to date
% in the context of solar flare effectively retained their original
% stochastic driver

Evidently, the form of deterministic driving embodied in Equation~(\ref{eq:detdrive})
precludes using the initial condition $A_{i,j}^0=0$. In what follows we typically
use a lattice having reached the SOC state via the original LH model, and
then switch to deterministic driving. This will typically lead to a transient
phase during which the lattice reajusts to the new driving scheme, and/or
other altered model components, as discussed presently. This transient
phase is usually accompanied by a gradual change in the total lattice
energy, so that monitoring the latter is the best way to ascertain
the recovery to a statistically stationary avalanching state.

\subsection{Stochastic Redistribution Rules\label{ssec:storedst}}

Directly incorporating the deterministic driving rule given
by Equation~(\ref{eq:detdrive})
within the Lu and Hamilton model yields a deterministic cellular
automaton characterized by a periodic cycle of energy loading/unloading,
all energy release events having the same magnitude. To recover
SOC-like behavior, some stochastic element need be reintroduced
somewhere in the model.

\subsubsection{Random Redistribution\label{sssec:rr}}

Consider the following,
plausible stochastic variation on the Lu and Hamilton
redistribution rules:
the same amount of nodal variable [$Z$] is always removed
from any unstable site, following the \citet{Lu:1993cy}
prescription, but it is distributed randomly, rather than equally,
amongst its four nearest neighbours:
\begin{eqnarray}
  \label{eq:redis3}
  A_{i,j}^{n+1}&=& A_{i,j}^n-\frac{4}{5}Z~, \\
  \label{eq:redis4}
  A_{i\pm 1,j\pm 1}^{n+1}&=& A_{i\pm 1,j\pm 1}^n+{4\over 5}\left({r_k\over R}\right)Z~,
\end{eqnarray}
where the $r_k$ ($k=1,...,4)$ are random deviates uniformly distributed
in $[0,1]$, and $R=\sum r_k$.
This scheme will be referred
to as {\it random redistribution}. Note that here that some particular
random sequences and node configurations may lead to negative energy
release from the redistribution process. We avoid such unphysical
events by permuting the four random numbers over the neighbouring
nodes and then sometimes generate four new random numbers
until a positive energy is released by the avalanche. % (empirically, at
%least one of the permutations always satisfies this condition).

\subsubsection{Random Extraction\label{sssec:re}}

Another means of introducing stochasticity is
to retain equal redistribution to nearest neighbours, but let
the quantity $Z$ extracted from the unstable node
be drawn from a bounded distribution
of uniform random deviates
\begin{equation}
  \label{eq:redis5}
  Z\in [|\Delta A_{i,j}^n|-Z_c, Z_c]~,
\end{equation}
while retaining Equations~(\ref{eq:redis1}) -- (\ref{eq:redis2}) for the
redistribution {\it per se}.
The lower bound in Equation~(\ref{eq:redis5}) ensures that the redistribution will,
at worst, restore marginal stability, while the upper bound corresponds
to the setting of the Lu and Hamilton model. This scheme will be referred
to as {\it random extraction}, and, defined in the above manner,
involves no adjustable parameters once $Z_c$ has been specified.

\subsubsection{Nonconservative Redistribution\label{sssec:nc}}

A third approach is to introduce {\it non-conservative} redistribution
rules. In all rules discussed so far -- including the original Lu and Hamilton
rules -- redistribution is conservative, in that whatever quantity of
$A$ being extracted from an unstable node ends up in the nearest neighbours,
whether in equal (Equations~(\ref{eq:redis1}) -- (\ref{eq:redis2}))
or unequal (Equations~(\ref{eq:redis3}) -- (\ref{eq:redis4})) proportions.
This conservation property is basically inspired by the sandpile
analogy, where avalanches redistribute sand grains without
creating or destroying any.
With a nodal variable associated
with a single component of the magnetic vector potential, there exist no
physical requirement for conservative redistribution of $A$; the
important physical
requirement, namely $\nabla\cdot\vec{B}=0$, is already taken care
of under the current interpretive scheme, see Equation~(\ref{eq:bdef}).

A nonconservative version of the Lu and Hamilton redistribution
rules can be defined as follows:
\begin{equation}
  \label{eq:redis8}
  A_{i,j}^{n+1}= A_{i,j}^n-\frac{4}{5}\,Z~,
\end{equation}
\begin{equation}
  \label{eq:redis9}
  A_{i\pm 1,j\pm 1}^{n+1}= A_{i\pm 1,j\pm 1}^n+{r_0\over 5} Z~,
\end{equation}
where $r_0\in[D_{\rm nc},1]$ is again extracted from a uniform distribution
of random deviates with a lower bound [$D_{\rm nc}$ ($<1$)], such that
$1-D_{\rm nc}$ is the
fraction of the redistributed quantity $Z$ that is lost rather
than redistributed. This rule thus involves one free parameter,
namely the conservation parameter [$D_{\rm nc}\in [0,1[$].
A nonconservative model of this type,
using fully deterministic driving, redistribution, and stability criteria, has been
developed by \citet{Liu:2006cp,VallieresNollet:2010ed} in the context of magnetospheric
substorms, and has been shown to produce avalanches with SOC-like power-law
distribution for event sizes.

\subsection{Stochastic Threshold\label{ssec:stothresh}}

Another interesting possibility is to introduce
stochasticity at the level of the stability threshold.
This can be achieved, for example, by replacing
the deterministic threshold rule given by Equation (\ref{eq:stab1}) by
extracting a value $Z_c$ anew at each node of each
temporal iteration from a sharply peaked normal
distribution centered on some mean value ${\bar Z_c}$.
An important parameter in such a scheme
is the width at half-maximum [$\sigma$] of the Gaussian distribution;
if $\sigma$ is very small (in the sense that
$\sigma/{\bar Z_c}\ll 1$,
one expects behavior similar to the conventional,
fixed threshold stability rule, while a very wide Gaussian
turns the stability threshold into a random triggering process
driven by noise.

\citet{Lu:1993cy} experimented with
similar random variations of the threshold rules
in the context of their stochastically driven model,
but report no significant variations of the results
as compared to their deterministic threshold rule.
As will become apparent further below,
in the context of a deterministically driven model,
however, interesting effects are produced
in the regime $\sigma/{\bar Z_c}\ll 1$.\\

To sum up, we are displacing the stochastic element
in the Lu and Hamilton model from the forcing rule to
either the redistribution rule or the threshold rule.
In the original Lu and Hamilton model, avalanche dynamics
is fully deterministic but the driving is stochastic.
The model variations outline above move the stochastic
element to the avalanche dynamics, in the context of
a spatially global and fully deterministic driving mechanism
that can be interpreted plausibly as global twisting
of a coronal loop.

\section{Results\label{sec:results}}

In principle, fully deterministic avalanche models, \textit{e.g.} the LH model with
the deterministic driving defined by Equation~(\ref{eq:detdrive}) and a smooth
initial condition, lead to
a regular cycle of energy loading/unloading, with avalanches
equally spaced in time and all liberating the same amount of
energy. The challenge in reintroducing stochasticity in the
stability and/or redistribution rules is to break this deterministic
loading/unloading cycle \citep[in this context, see also][and
references therein]{Wheatland:1998gx}.

In the preceding section we have introduced various classes
of redistribution rules based on a number of stochastic elements,
as well as a stochastic threshold rule. A large number of
distinct avalanche models can be constructed by picking and
combining this or that model element. One could, for instance,
build a fixed-threshold model using nonconservative redistribution rules
(Section \ref{sssec:nc})
with equal distribution to nearest neighbors; or a
conservative rule with random distribution to nearest-neighbors
(Section \ref{sssec:rr}),
in conjunction with a stochastic threshold rule (Section \ref{ssec:stothresh}).
In addition, each such model may involve adjustable parameters
(\textit{e.g.} the width [$\sigma$] of the Gaussian distribution from which the
threshold values are extracted). Although we have explored a vast
portion of the relevant model space, in what follows we focus
on a relatively small subset of such models, chosen so as
to illustrate the range of behaviours possible in deterministically driven
avalanche models. In order to expedite this exploration of parameter
space, most model runs were carried out on a small grid
($48^{2}$), but we have also recomputed a number of runs
on larger grids (up to $384^{2}$) to ensure that the results
reported upon below are robust with respect to lattice size.

Exploration of model space quickly reveals that an important
behavioral discriminant is the conservative property (or lack
thereof) of the redistribution rule. Consequently we first
discuss results for a variety of conservative models
(Section \ref{ssec:rescons}), and turn subsequently
to their non-conservative counterparts
(Section \ref{ssec:resncons}). We finally considers (Section \ref{ssec:fss}) the
origin of the break of finite size scaling that takes place in our
models. 

\subsection{Conservative Redistribution\label{ssec:rescons}}

Table \ref{tab:cons} lists the defining rules and parameter values of a 
set of representative conservative avalanche models, whose
properties will be discussed in the remainder of this section.
Here and in all that follows, unless noted otherwise
all models are run on a $48^{2}$ lattice, 
use the driving scheme described by Equation~(\ref{eq:detdrive}), and
a stability threshold $Z_c=1$
(or ${\bar Z_c}=1$ for models with a stochastic threshold).
A ``0'' entry in the second column indicates
a deterministic threshold rule. The entry ``random'' in the third and
fourth columns indicates use of Equations (\ref{eq:redis5}) and
(\ref{eq:redis3}) -- (\ref{eq:redis4}), respectively.
The last line in the Table lists the ``ingredients'' of the LH model, for
comparison. %Note that we also considered a non-stochastic model C0 for completeness.

\begin{table}
\begin{center}
\begin{tabular}{lrlllll}
\toprule
{} & $\sigma/Z_c$ & Extraction & Redistribution &        $\alpha_E$ &        $\alpha_P$ &        $\alpha_T$ \\
\midrule
C1 &          0.0 &     Random &  $4\times 1/5$ &  $1.43 \pm 0.044$ &  $1.49 \pm 0.243$ &  $1.94 \pm 0.136$ \\
C2 &         0.01 &      $Z_c$ &  $4\times 1/5$ &  $1.47 \pm 0.008$ &  $1.84 \pm 0.019$ &  $1.92 \pm 0.023$ \\
C3 &          0.0 &      $Z_c$ &         Random &  $1.29 \pm 0.026$ &  $1.42 \pm 0.045$ &  $1.38 \pm 0.022$ \\
C4 &         0.01 &     Random &  $4\times 1/5$ &  $1.40 \pm 0.013$ &  $1.56 \pm 0.087$ &  $1.77 \pm 0.033$ \\
C5 &         0.01 &      $Z_c$ &         Random &  $1.29 \pm 0.015$ &  $1.47 \pm 0.037$ &  $1.38 \pm 0.011$ \\
C6 &          0.0 &     Random &         Random &  $1.22 \pm 0.009$ &  $1.12 \pm 0.037$ &  $1.47 \pm 0.008$ \\
C7 &         0.01 &     Random &         Random &  $1.19 \pm 0.015$ &  $1.08 \pm 0.034$ &  $1.42 \pm 0.004$ \\
LH &          0.0 &      $Z_c$ &  $4\times 1/5$ &  $1.48 \pm 0.005$ &  $1.72 \pm 0.143$ &  $1.63 \pm 0.023$ \\
\bottomrule
\end{tabular}
\end{center}
\caption{Conservative avalanche models, with a driving
  parameter $\varepsilon =10^{-6}$. Error bars
  were obtained with ten different random number sequences.}
\label{tab:cons}
\end{table}

\begin{figure}
  \centering
  \includegraphics[width=\linewidth]{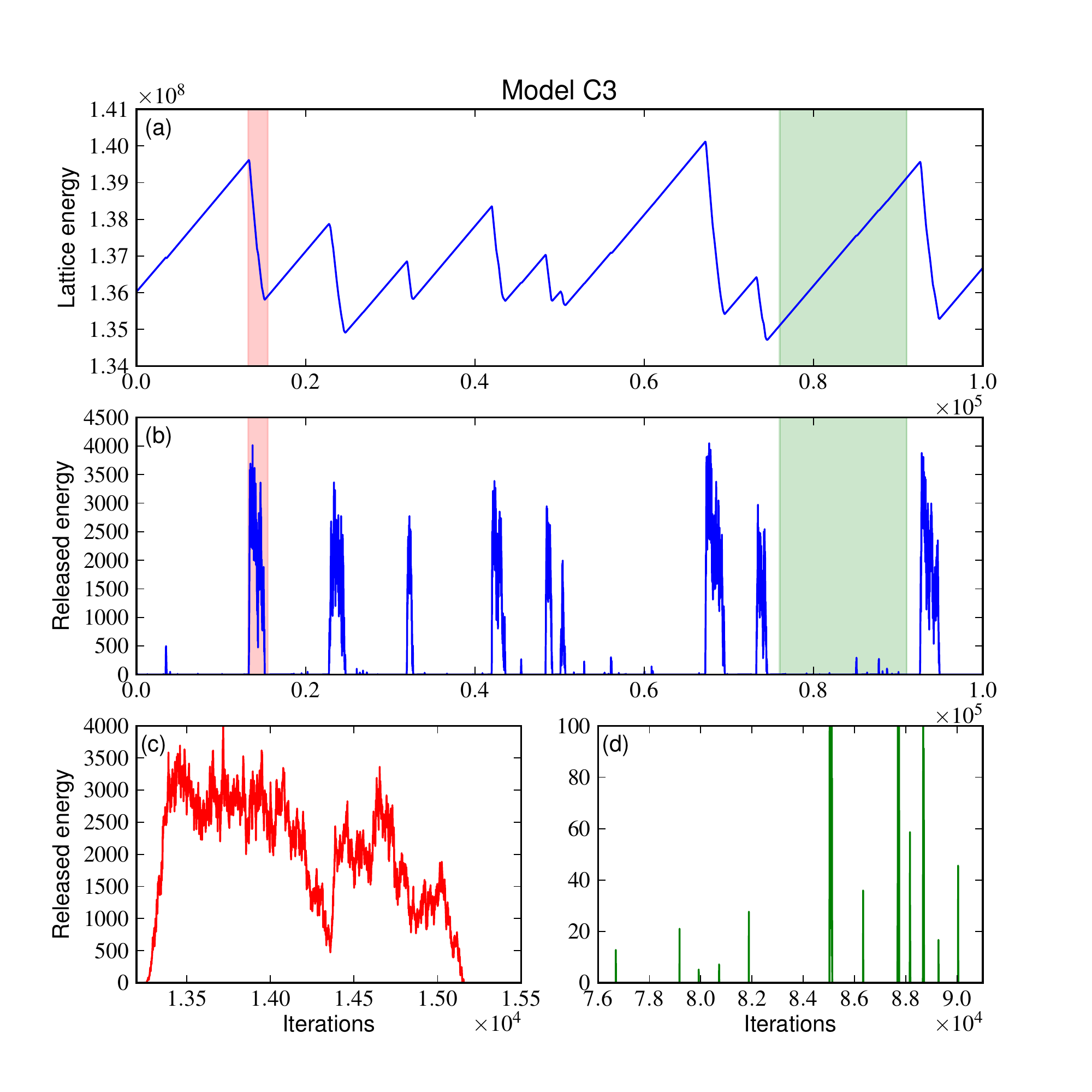}
  \caption{Sample time series for avalanche model C3. (a)
    Temporal evolution of total lattice energy. (b) Energy released by
    avalanches. There are two populations
    of avalanches here: quasiperiodic large avalanches (c), in between which
    occurs a population of smaller avalanches spanning a wide range of scales
    in size (d). Energy is measured in the elementary unit $e_0$ given by
    Equation~(\ref{eq:eunit}).}
  \label{fig:tscons}
\end{figure}

\begin{figure}
  \centering
  \includegraphics[width=\linewidth]{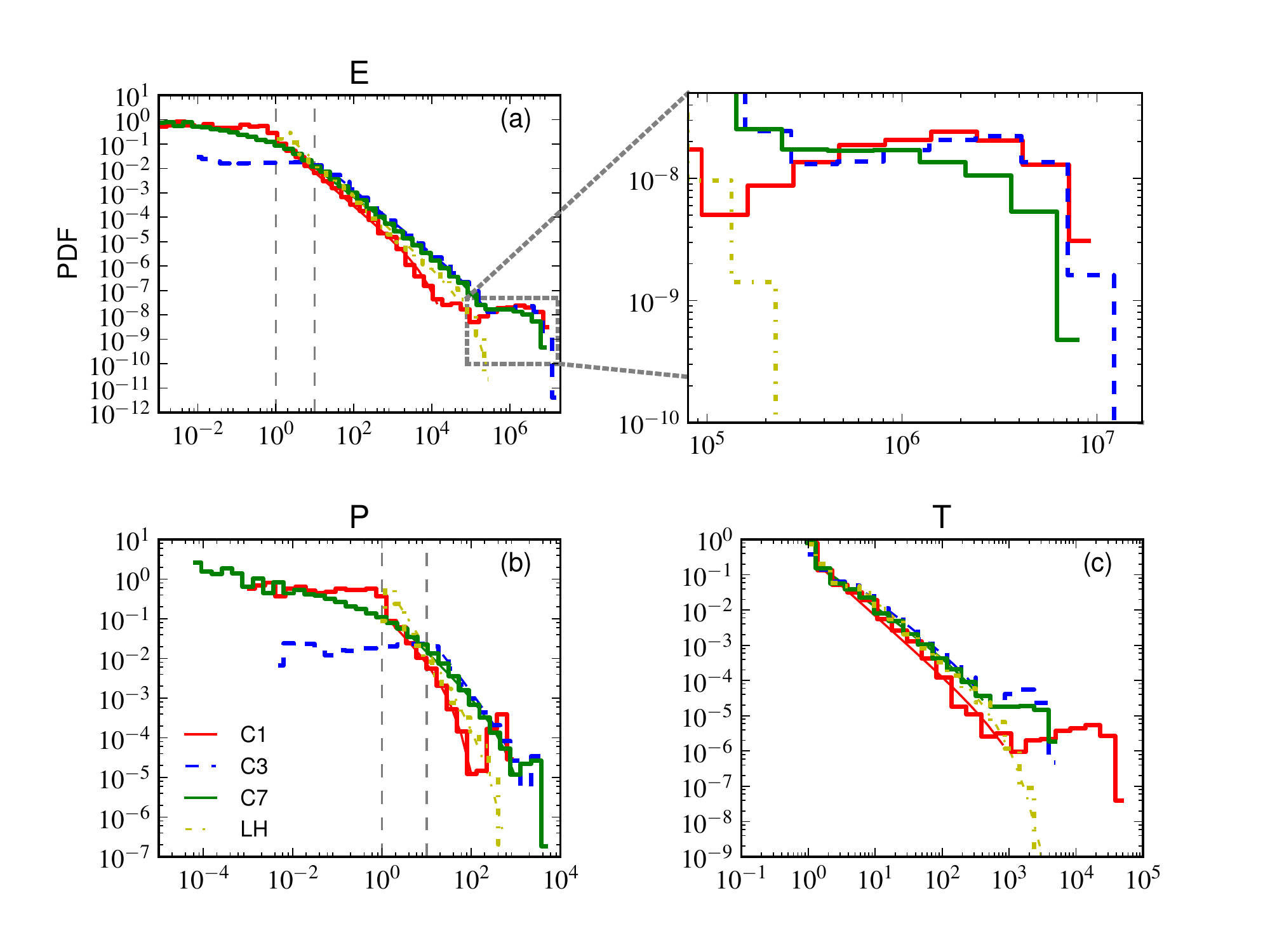}
  \caption{Statistical properties of avalanches in models C1, C3, C7,
    and LH, as
    labeled. The various panels show
    the probability distribution
    functions of (a) total energy released by avalanches [$E$],
    (b) peak energy release [$P$], and (c) duration [$T$]. A close-up
    of the high-$E$ end of the total energy PDF is also shown.
    The straight lines are guides to the eye, showing
    the slopes associated with the mid-sized avalanches
    of the distributions.
    All PDFs
    are constructed from $10^8$ iterations runs, initialized
    from prior runs of the same models having already reached
    a statistically stationary state. The vertical dashed lines label
    the transition from small to power-law avalanches (see discussion
    a the end of section \ref{sec:results}).}
  \label{fig:pdfcons}
\end{figure}

Figure~\ref{fig:tscons} shows a short segment of the time series for
total lattice energy (a) and energy release (b) in model
C3. This is a model using deterministic extraction, random redistribution,
a fixed stability threshold $Z_c$,
and a deterministic driving parameter $\varepsilon=10^{-6}$.
The lattice energy time series exhibits a pattern of quasiperiodic
energy loading and unloading. As can be seen in 
Figure~\ref{fig:tscons}b, the unloading takes place via a sequence of large avalanches
of roughly similar amplitudes and durations taking place at more or less regular
time intervals. Comparison between panels c and a reveals that 
a small but significant fraction ($\lesssim 1 \%$)
of the total lattice energy is dissipated
by any one of
these large avalanches, in between which lattice energy grows
almost steadily in response to driving.
Careful scrutiny of Figure~\ref{fig:tscons}b also reveals a population of
much smaller avalanches interspersed between these large ones.
Figure~\ref{fig:tscons}d is a zoom with an
expanded vertical scale, spanning an
interval between two large avalanches and showing
the smaller ones in better detail.
Unlike the quasiperiodic large avalanches, these are seen to span
a wide range of amplitudes, and show no hint of any (quasi)periodicities.
Even the largest avalanches in this second population only release
a minuscule fraction of the lattice energy, and are hardly visible
during the energy loading rising phases in Figure~\ref{fig:tscons}a.
All of this suggests the presence of (at least) two dynamically distinct
avalanche populations. This impression is confirmed upon
computing the probability distribution functions (hereafter PDF)
for total energy release, \textit{i.e.} the probability of
finding an avalanche having released a total energy within
the interval $[E,E+{\rm d}E]$. This PDF, as well
as the corresponding PDF for peak energy release [$P$]
and avalanche duration [$T$], are
shown on Figure \ref{fig:pdfcons} 
for a set of models.
The bulk part of all three PDFs is well
fitted by a power-law for all models, except at the high ends of the
distributions. A marked excess of large, long duration
events shows up
as a bump -- well fitted by a Gaussian --, and for cases with random
redistribution/extraction a 
significant depletion for very low avalanches occurs (a detailed discussion
on to this particular population is given at the end of this section). It is readily
verified that the bump at the lower end of the PDF corresponds
to the large avalanches, while the power-law portion of the PDF
is made up of all avalanches taking place in between, as in
Figure~\ref{fig:tscons}d.

\begin{figure}
  \centering
  \includegraphics[width=\linewidth]{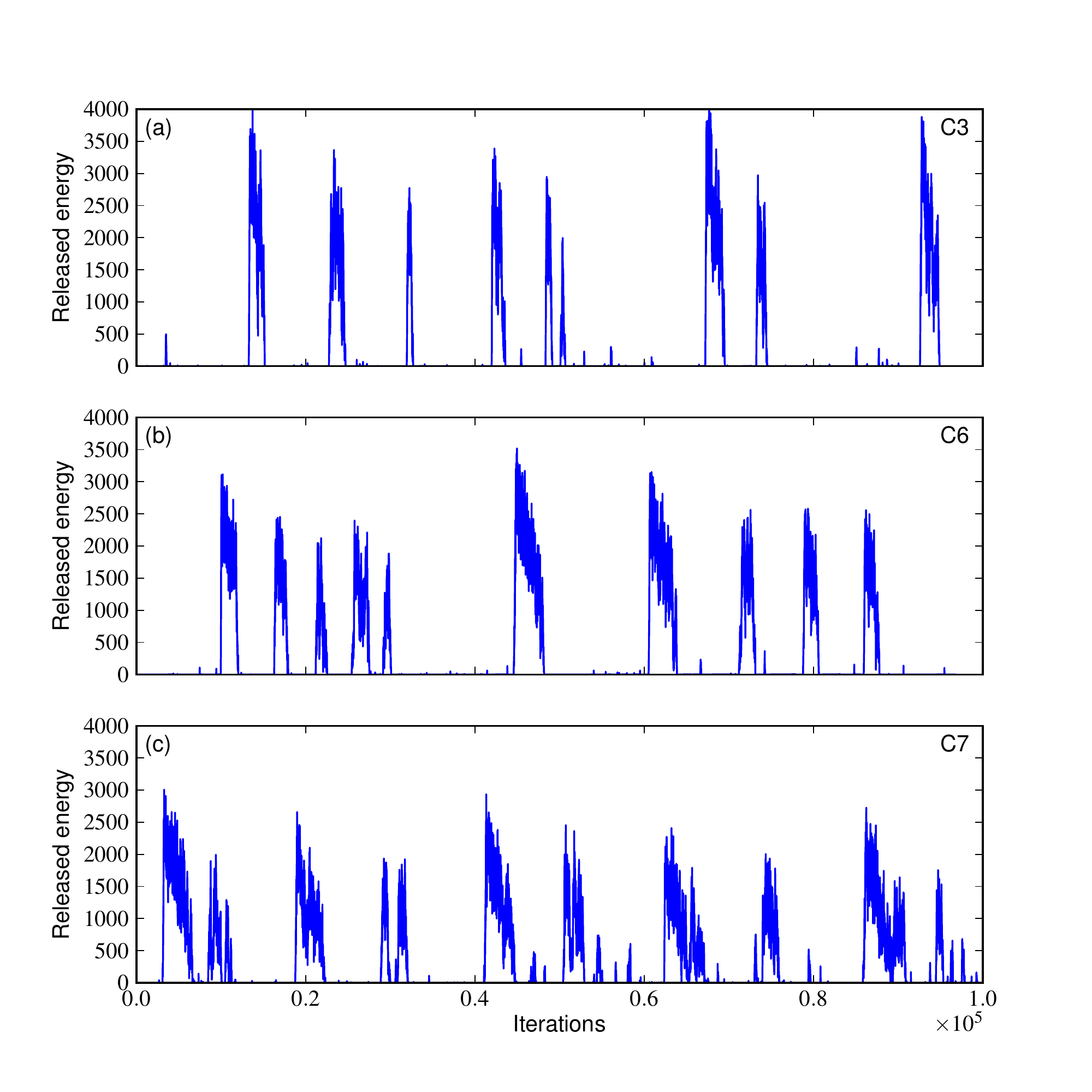}
  \caption{Extracts of energy release time series in 
    models C3, C6, and C7 (panels a through c). This
    represents a sequence of increasing stochasticity
    built into the avalanche models (cf. Table \ref{tab:cons}).
    All models use the same deterministic driving rate,
    namely $\varepsilon=10^{-6}$ in Equation~(\ref{eq:detdrive}).}
  \label{fig:3ts}
\end{figure}

Evidently, in the case of model C3, the stochasticity
introduced at the level of redistribution
is insufficient to break the quasiperiodic pattern of
energy loading and unloading, as manifested by the population
of large avalanches. Indeed, this pattern proves surprisingly difficult
to break. Figure \ref{fig:3ts} shows $10^5$ iterations
segments of energy release
time series extracted from a sequence of three avalanche models,
all operating at the same deterministic driving rate $\varepsilon=10^{-6}$,
but
including more and more stochastic components (specifically, models
C3, C6, and C7 in Table \ref{tab:cons}). Comparing the three panels reveals
immediately that increasing stochasticity gradually breaks
the quasiperiodicity of large avalanches, while blurring
the energy contrast between the populations of large and small
avalanches, mostly by reducing the size of the large loading/unloading
avalanches. Corresponding PDFs for the integrated energy release [$E$],
peak energy release [$P$] and avalanche duration [$T$] are also shown
on Figure~\ref{fig:pdfcons}. Even in the case of the strongly stochastic
model C7, these PDFs
(solid green histograms in Figure~\ref{fig:pdfcons})
still show an excess of large events as compared to a pure,
scale-free power-law distribution. % At higher driving rates,
% the population of large avalanches can become completely disconnected
% from the small avalanches in the PDFs for $E$, $P$ and $T$.
% for example, in model C3 with $\varepsilon=10^{-5}$, this gap
% in the PDF for total energy spans over an order of magnitude.

Further insight into this dichotomy in avalanching behavior is
obtained by computing a waiting time distribution (WTD)
separately for each class of avalanches.
The waiting time [$\Delta T$] associated with the $k{\rm th}$
avalanche is defined as the time elapsed
since the end of the previous % large
avalanche and the beginning of that under consideration.
The distinction between ``large'' and ``small'' avalanches is
made on the basis of the PDF for peak energy release, which typically
shows most clearly the bump associated with the population of large 
avalanches; as the PDF is scanned from the low end of the size distribution
towards the high end,
a point is reached where the PDF shows a local minimum
(\textit{e.g.} at $P/e_0\approx 2\times 10^2$ for model C1 in
  Figure~\ref{fig:pdfcons}b). All avalanches 
located right of this point are considered to belong to the
population of ``large'' avalanches, even though they likely contain a few
of the largest avalanches belonging to the second population of
``small'' avalanches, but there is simply no way to reliably distinguish these.

\begin{figure}
  \centering
% %\includegraphics[scale=0.8]{figuregauss1.eps}
  %\includegraphics[width=\linewidth]{Case_C3_eps}
  \includegraphics[width=\linewidth]{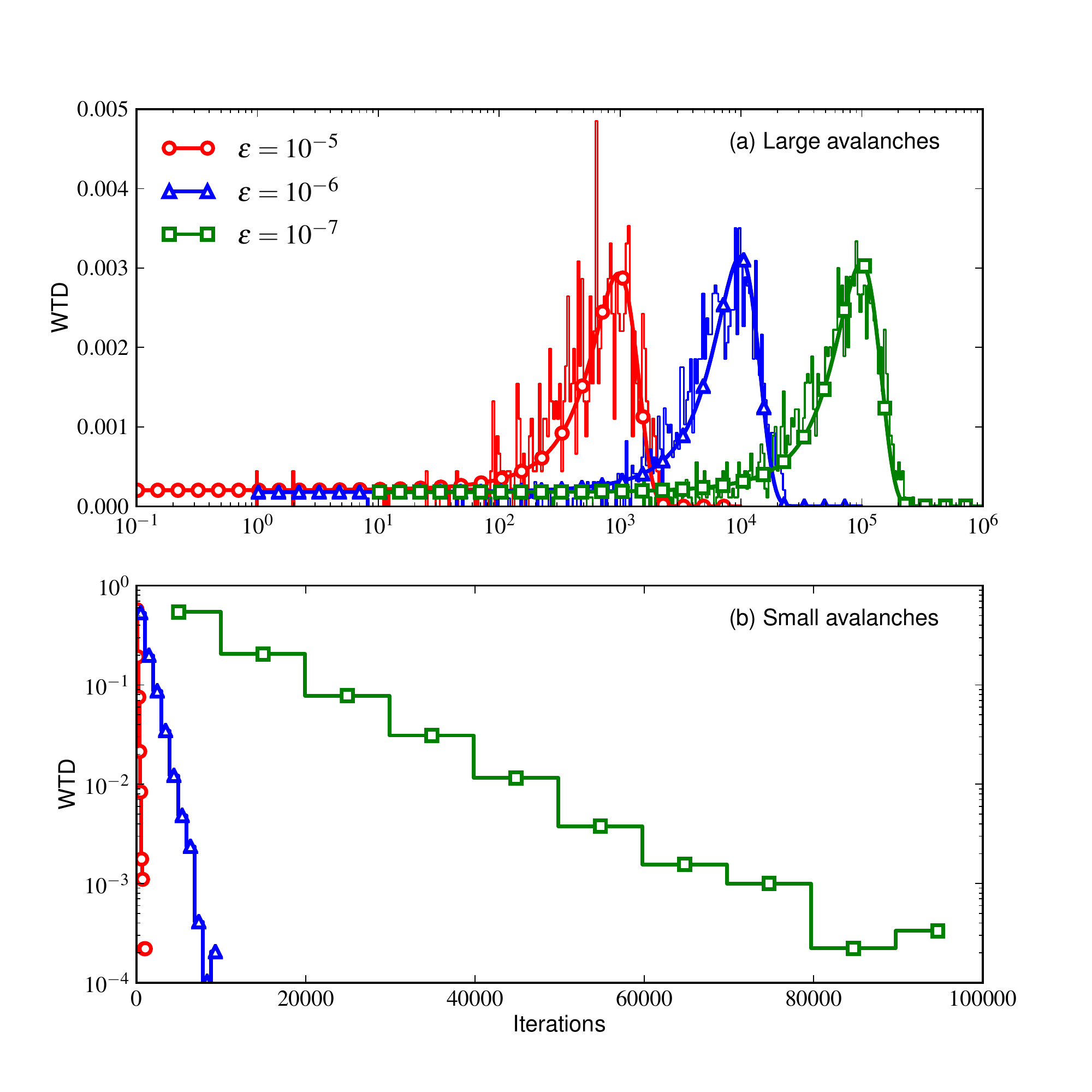}
  \caption{Waiting-time distributions for the populations of (a) large and
    (b) small avalanches, for three C1-type models with
    deterministic driving rates of $\varepsilon=10^{-7}$,
    $10^{-6}$, and $10^{-5}$, as labeled. The WTDs
    for large avalanches are well fitted with a Gaussian (solid lines
    on panel a),
    while those characterizing small avalanches have an exponential form.}
  \label{fig:wtdcons}
\end{figure}

The WTDs for large and small avalanches are plotted
in Figure~\ref{fig:wtdcons}a -- b, here for C1-type models with
driving rates $\varepsilon=10^{-7}$, 
$10^{-6}$, and $10^{-5}$.
Large avalanches
have a well-defined mean wait time and their WTD is well
fitted by a Gaussian, while the population
of smaller avalanches has an exponential WTD, indicative of
triggering by
a stationary random process. Increasing the driving
reduces the mean wait time for large avalanches in direct proportion,
while steepening the exponential WTD for the population of
small avalanches. This indicates that the large avalanches
are the (perturbed) signature of the energy loading/unloading
process that would otherwise characterize a fully deterministic
model.
For the population of small avalanches, increasing the driving rate
merely steepens the WTD while maintaining its exponential form.
This is the same behavior observed in the LH model when the
mean amplitude of the spatially random increments remains constant
in time and space \citep[on the WTD in the LH model, see also][]{Norman:2001co,Wheatland:2000if}.

\begin{figure}
  \centering
  \includegraphics[width=\linewidth]{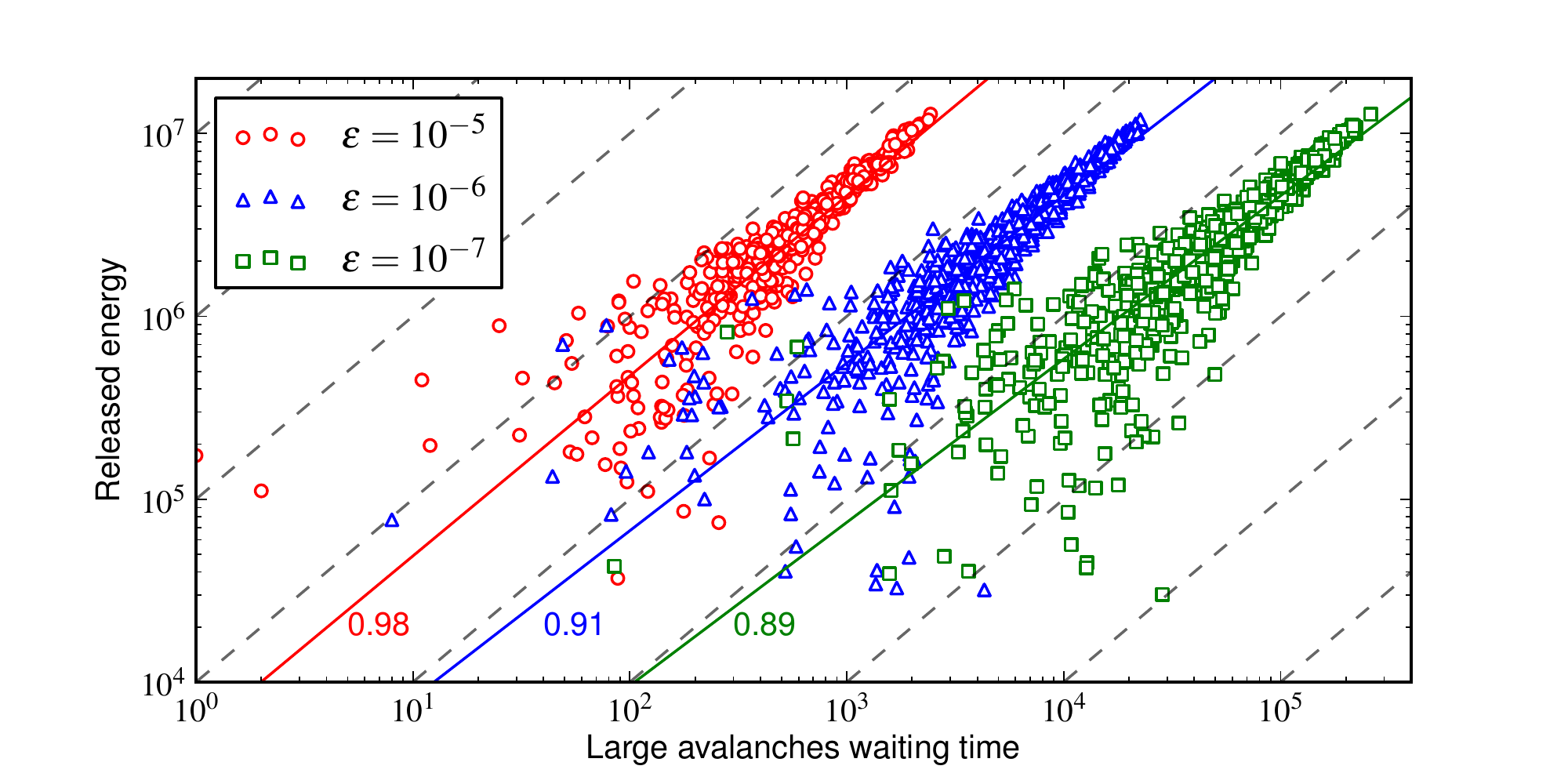}
  \caption{Correlation between the energy of large
    avalanches, and the waiting time elapsed since the end
    of the previous large avalanche. Scatter plots are shown for
    C1-type models with varying driving rates [$\varepsilon$],
    as indicated. 
    %The diamonds indicate the bivariate mean of the distributions.
    The dashed lines are a guide to the eye
    corresponding to a logarithmic slope of unity, indicating
    linear proportionality between avalanche energy and wait
    time, as expected in simple
    energy loading/unloading models (see text).}
  \label{fig:Ewtd}
\end{figure}

This is confirmed upon constructing scatter plots between
the energy of large avalanches and the wait time elapsed since
the end of the previous large avalanche, as done on
Figure \ref{fig:Ewtd}, for a sequence of C1-type models
spanning four decades in driving rates: $\varepsilon=10^{-7}$,
$10^{-6}$, and $10^{-5}$. For a simple, fully deterministic
load/unload model \citep[\textit{e.g.} of the type originally considered
by][]{Rosner:1978fk}, avalanches are periodic and all liberate the
same amount of energy; all points would then coincide.
For a loading/unloading model including a stochastic trigger,
one would expect a strong positive correlation between avalanche
energy and wait time. For the spatially extended models considered here,
individual avalanches are scattered in a cloud, whose extent in the
$[E,\Delta T]$ plane increases with increasing degree of stochasticity in the
avalanche model. This occurs because the unfolding of a large
energy-unloading avalanche is affected by the spatial distribution
of nodal variable values, itself influenced by the stochastic
components introduced in the redistribution or threshold rules.
Nonetheless, linear proportionality between energy and wait time
is clearly apparent for the higher driving rates, but degrades
towards lower $\varepsilon$-values; the linear correlation coefficients
are $r=0.98$, $0.91$, and $0.89$ for $\varepsilon$ falling from
$10^{-5}$ to $10^{-7}$, in decadal steps (see Figure \ref{fig:Ewtd}).%  The slight decrease in $r$
% in going from $\varepsilon=10^{-5}$ up to $10^{-4}$ stems from the
% difficulty to reliably distinguish the two populations of avalanches
% at this high driving rate.

%

\begin{figure}
  \centering
  \includegraphics[width=0.5\linewidth]{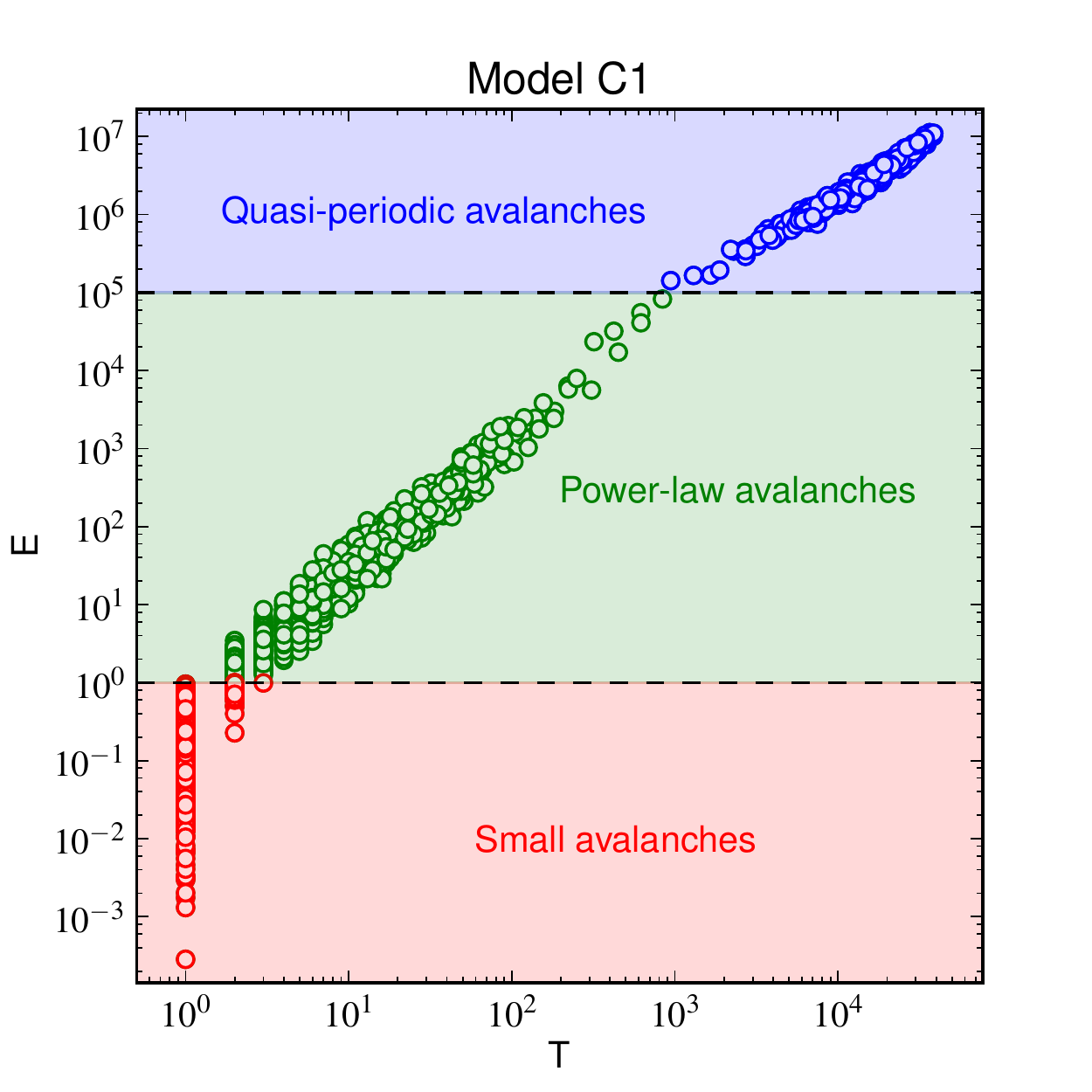}
  \caption{[$E$] \textit{vs} [$T$] in model C1. The three populations of
    avalanches (small, mid-size power law and large quasi-periodic)
    are well identified in three energy regions. }
  \label{fig:E_vs_T}
\end{figure}

The models embedding a stochastic process in their
redistribution or in their extraction rule (see Table \ref{tab:cons}) all
exhibit a particular population of small avalanches that is not
well-fitted by a power-law (\textit{e.g.} Figure \ref{fig:pdfcons}a -- b).
These avalanches have an energy of the order of and lower than the unit
energy [$e_{0}$]. In the case of the LH model, one-iteration -- and
one-node -- avalanches always
release an energy of $e_{0}$ whereas in the case of random
redistribution or extraction, such avalanches can release a large
range of different energies. This population of small avalanches -- of
few iterations -- appears to deviate from the classical SOC state of the LH
model. They can be easily identified in Fig.
\ref{fig:E_vs_T} where we scatter avalanches as a function of their
duration [$T$] and released energy [$E$] for model C1. The alteration of
very small avalanches can be different from one model to the other: model
C1 shows a deviation for $E < e_{0}$ while model C3 is affected up to
$E\approx 10\,e_{0}$. Despite the existence of scale-dependent populations at both
ends of the avalanche distribution, the total energies of mid-size avalanches are still very
well fitted by a power-law over more than four orders of magnitudes
(more than two orders of magnitude for the peak energy).

The last three columns of Table \ref{tab:cons} list the power-law indices obtained
from least-square fits to the portions of the PDFs of total energy,
peak energy release, and duration for the populations of mid-size
avalanches in the corresponding simulations (the error bars were
obtained with 10 runs using different random numbers sequences).
These power-law indices are rather similar to
those extracted from the 2D scalar version of the LH stochastically driven
avalanche model (last line in Table \ref{tab:cons}). Significant differences are
observed nonetheless in going from model to model, suggesting that
global statistics of event sizes are dependent to some extent
on model ``ingredients'' and parameters in these conservative models.

We have explored a wide variety of conservative models, often combining
into the same model multiple stochastic elements (see models C1 through C7
in Table \ref{tab:cons}).
Remarkably, none of these models succeeds in completely breaking
the loading/unloading cycle. While producing scale-free avalanche size
distributions spanning many decades in energy, all models also end up producing
a population of very large quasi-periodic events with a
more-or-less well-defined mean size. Even model
C7, combining a fairly broad random stability threshold, random extraction, and
random redistribution, ends up doing so. This property is truly
robust, and holds over a wide range of driving rates.
In contrast, the PDFs for flare sizes reconstructed from solar UV or soft X-ray
data shows no such population, and is instead well represented
by a single power-law \citep[see][and references
therein]{Dennis:1985ev,Lu:1993cy,Aschwanden:2000gc,Aschwanden:2002jy}. As
detailed in what immediately follows,
the use of nonconservative redistribution rules can alleviate this problem
over a certain range in parameter space, for all classes of
deterministically driven
models. 

\subsection{Nonconservative Redistribution\label{ssec:resncons}}

We consider once again a representative set of deterministically driven
avalanche model using non-conservative redistribution, as described in
Section \ref{sssec:nc}, and various combination of other modeling
ingredients. Table \ref{tab:ncons}
lists the characteristics and properties of the models that are the
focus of the foregoing discussion. 
The general format is similar to Table \ref{tab:cons}, and the models
were run with a non-conservation parameter $D_{nc}=0.1$.
Recall that the actual fraction removed is drawn from
a sequence of uniform deviates spanning $[D_{nc},1]$, so that the {\it average}
nodal dissipation is the median of this interval. We essentially
discuss in this section model NC0, the other models follow naturally
the properties described in this and previous sections.

\begin{table}
\begin{center}
\begin{tabular}{lrlllll}
\toprule
{} & $\sigma/Z_c$ & Extraction & Redistribution &        $\alpha_E$ &        $\alpha_P$ &        $\alpha_T$ \\
\midrule
NC0 &          0.0 &      $Z_c$ &  $4\times 1/5$ &  $1.17 \pm 0.005$ &  $1.25 \pm 0.097$ &  $1.15 \pm 0.015$ \\
NC1 &          0.0 &     Random &  $4\times 1/5$ &  $1.12 \pm 0.015$ &  $1.08 \pm 0.037$ &  $1.22 \pm 0.007$ \\
NC2 &         0.01 &      $Z_c$ &  $4\times 1/5$ &  $1.20 \pm 0.015$ &  $1.23 \pm 0.080$ &  $1.16 \pm 0.066$ \\
NC3 &          0.0 &      $Z_c$ &         Random &  $1.19 \pm 0.007$ &  $1.25 \pm 0.116$ &  $1.15 \pm 0.009$ \\
NC4 &         0.01 &     Random &  $4\times 1/5$ &  $1.12 \pm 0.014$ &  $1.04 \pm 0.042$ &  $1.31 \pm 0.098$ \\
NC5 &         0.01 &      $Z_c$ &         Random &  $1.21 \pm 0.026$ &  $1.19 \pm 0.025$ &  $1.19 \pm 0.022$ \\
NC6 &          0.0 &     Random &         Random &  $1.12 \pm 0.021$ &  $1.08 \pm 0.036$ &  $1.22 \pm 0.010$ \\
NC7 &         0.01 &     Random &         Random &  $1.12 \pm 0.015$ &  $1.03 \pm 0.023$ &  $1.29 \pm 0.037$ \\
\bottomrule
\end{tabular}
\end{center}
\caption{Non-conservative avalanche models
  [$\varepsilon =10^{-6}$, $D_{\rm nc}=0.1$]. Error bars
  were obtained with ten different random-number sequences.}
\label{tab:ncons}
\end{table}

\begin{figure}
  \centering
% %\includegraphics[scale=0.8]{NC1ts.ps}
  %\includegraphics[scale=0.8]{figure6.eps}
  \includegraphics[width=\linewidth]{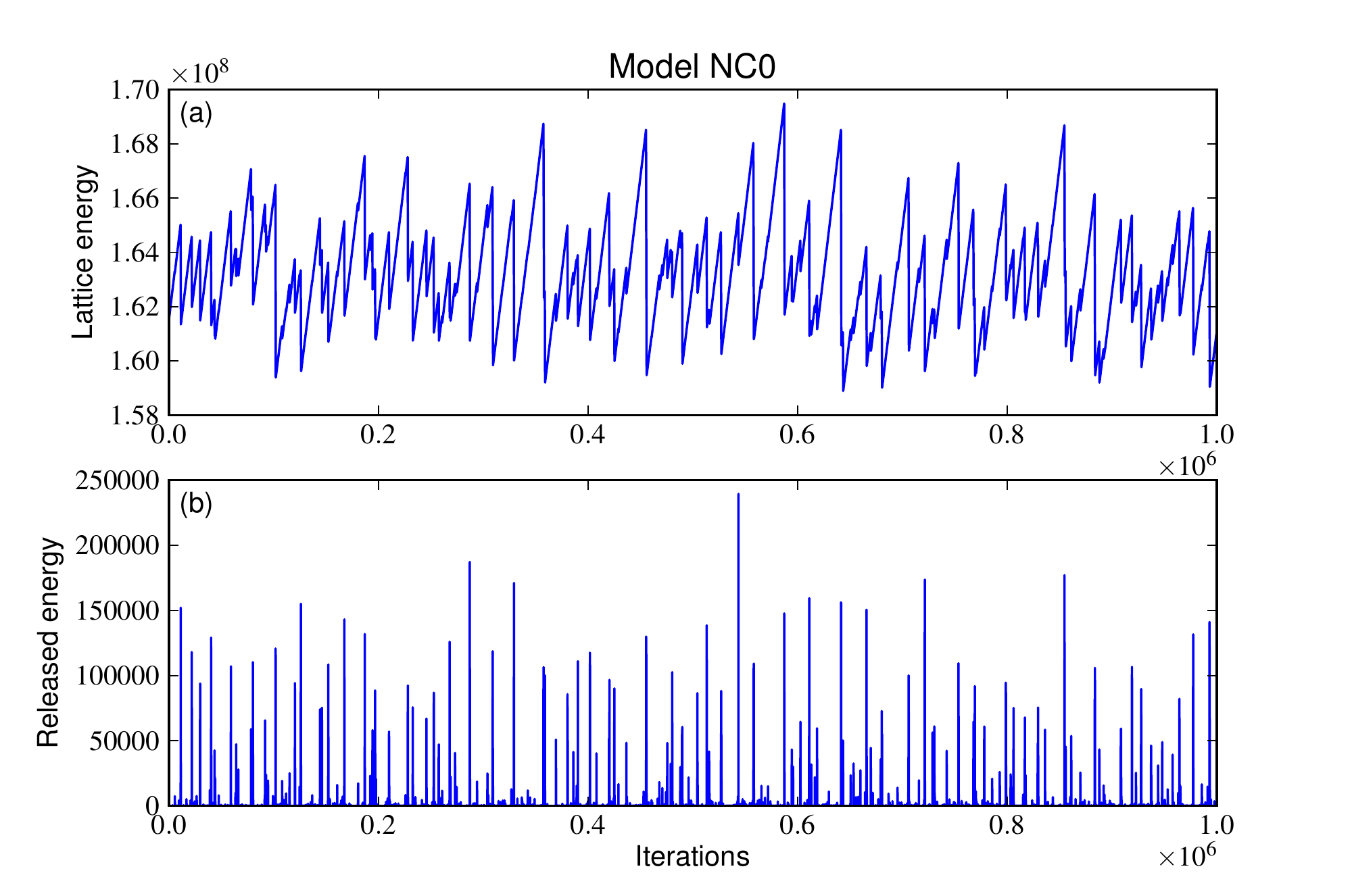}
  \caption{Sample time series for avalanche model NC0 ($D_{nc}=0.1$). Panel (a) shows
    the temporal evolution of total lattice energy, and part (b) the
    energy released by avalanches. There is now a single population
    of avalanches, without a characteristic event size or quasiperiodicity
    in event occurrence.}
  \label{fig:tsncons}
\end{figure}

Figure \ref{fig:tsncons}, similar in format to Figure~\ref{fig:tscons},
shows a $10^6$ iterations segment of the time series for
total lattice energy (a) and energy release (b) in model
NC0. This is now a nonconservative model using a driving rate
$\varepsilon=10^{-6}$, fixed threshold and
extraction, and equal redistribution to nearest-neighbors. The only
stochasticity introduced here is at the level of non-conservation, with
a random fraction between 0 and 90\% of the nodal variable value being lost
from each unstable node during redistribution.
Comparing to Figure~\ref{fig:tscons}a and b, one immediately notices the absence
of the very large quasi-periodic avalanches characterizing models with
conservative redistribution. The lattice-energy time series now has
the fractal-sawtooth form characteristic of the original
stochastically driven
LH model. The largest avalanches still only release a small fraction
of the total lattice energy (here about one percent at most, but much less on
larger lattices). The peak energy release of avalanches now spans over
three orders of magnitude even on the small $48^{2}$ lattice used here,
indicating that the larger avalanches span the whole lattice. No
hint of periodicities can be detected. These features suggest a SOC-like
state with a scale-free distribution of avalanche sizes.

\begin{figure}
  \centering
% %\includegraphics[scale=0.7]{PDF10-6b.ps}
  %\includegraphics[scale=0.7]{figure7.eps}
  \includegraphics[width=\linewidth]{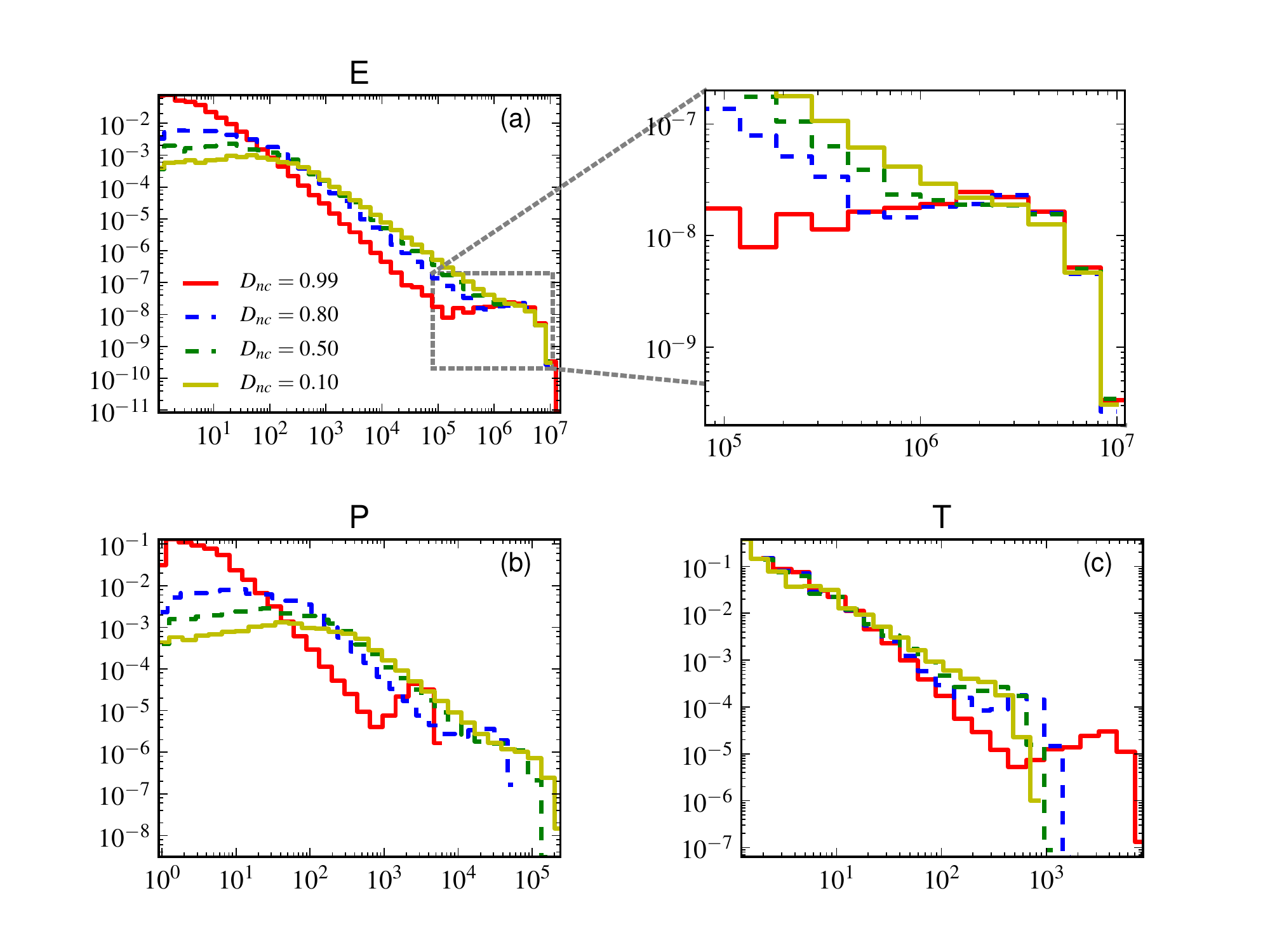}
  \caption{Probability distribution
    functions of total avalanche energy in a set of NC0-type
    model runs with fixed driving rate
    $\varepsilon=10^{-6}$ but decreasing conservation parameter
    $D_{\rm nc}$,
    as labeled. % $D_{\rm nc}=1$ corresponds naturally to model C0
    % (section \ref{ssec:rescons}).
    The layout is
    similar to Figure \ref{fig:pdfcons}: panel a, b and c show
    respectively $E$, $T$ and $P$.} 
  \label{fig:pdfncons}
\end{figure}

This impression is reinforced upon computing the PDFs for the size
of avalanches. Unless the conservation parameter
[$D$] tends towards unity (more about this shortly),
these PDFs now take the form of pure power laws for mid-size and large
avalanches (cf.
Figure \ref{fig:pdfncons}). Small avalanches
appear to systematically depart from a power law distribution in a
similar fashion to than observed for conservative models in Section
\ref{ssec:rescons}. The origin of the flattening is in fact the same:
small avalanches have now the possibility to release energies in a
much broader range than in the original LH model. This feature
notwithstanding, at small conservation parameter all of these
nonconservative models show a scale-free distribution for mid-size and large
avalanches in total avalanche energy, peak energy release and duration.

Models with moderate to high conservation ($D_{\rm nc}\to 1$)
do exhibit deviations from power-law behavior for
the largest events. The limit $D_{\rm nc}\to 1$ takes us back to conservative
models, so it is not surprising to see a second population of avalanches
appear. Figure \ref{fig:pdfncons},
shows a set of PDFs computed for a set of NC0-type model runs
with driving rate fixed at $\varepsilon=10^{-6}$ and decreasing values
of the dissipation parameter $D_{\rm nc}$, as labeled.
Recall (Table \ref{tab:ncons}) that except for the dissipation
mechanism, these models are otherwise fully deterministic.
The models at $D_{\rm nc}=0.1$ yield pure power
laws, but at $D_{\rm nc}=0.5$ an excess of large events
becomes apparent, which becomes quite pronounced at $D_{\rm nc}=0.99$.%  We
% observe very little variation of this behaviour with the driving rate
% $\varepsilon$ (not shown here).

\begin{figure}
  \centering
% %\includegraphics[scale=0.65]{alphancons.ps}
  %\includegraphics[scale=0.65]{figure9.eps}
  \includegraphics[width=\linewidth]{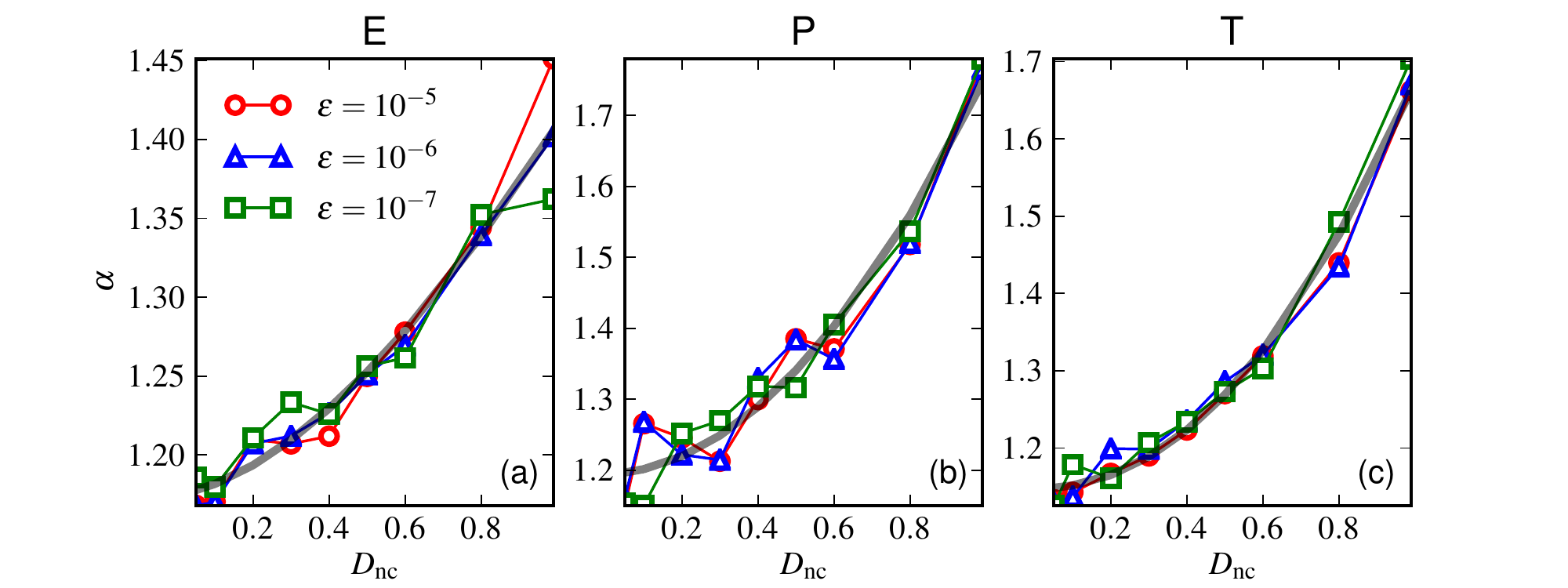}
 \caption{Variations of power-law indices for the PDFs of (a) total avalanche
    energy, (b) peak energy release and (c) duration, as a function of the conservation 
    parameter [$D_{\rm nc}$] and for various driving rates [$\varepsilon$], as labeled.
    The grey thick line corresponds to the power-law fit (see text).
    % Open squares indicate models where the event size distributions
    % are pure power laws, while solid dots indicate models where
    % a second population of large, quasiperiodic avalanches is present.
    % In such cases, only the power-law portion of the PDFs is
    % used to compute the $\alpha$'s. The gray horizontal bands indicate
    % the ranges in $\alpha$ spanned by model runs that show only
    % scale-free, power law PDFs in avalanche size.
    All these runs are for NC0-type models.}
  \label{fig:alphancons}
\end{figure}

 Figure \ref{fig:alphancons}
shows how the power-law indices for (a) avalanche energy, (b) peak release,
and (c) duration vary with driving rate [$\varepsilon$]
and conservation parameter [$D_{nc}$],
in the suite of NC0-type models runs.
% used to generate Fig.~\ref{fig:epsdplane}.
Each curve corresponds to a specific
driving rate [$\varepsilon$], as labeled. The power-law indexes show little to no
dependance to the driving rate -- provided it is sufficiently low --, as
expected. The general trend, namely steepening of the
PDFs with conservation parameter, also
characterizes the other types of nonconservative models listed in Table \ref{tab:ncons}.
A systematic dependence of the power-law indices with the conservation
parameter is observed in Figure \ref{fig:alphancons}. We fit the power
law indices [$\alpha$] as a function of a power law of $D_{nc}$ (gray
lines in Figure \ref{fig:alphancons}) and obtain:

\begin{eqnarray}
  \label{eq:fit_alphas}
  \alpha_{E} &=& 1.18 + \left(\frac{D_{\rm nc}}{2.48}\right)^{1.60} \, , \\
  \alpha_{P} &=& 1.20 + \left(\frac{D_{\rm nc}}{1.34}\right)^{2.00} \, ,\\
  \alpha_{T} &=& 1.15 + \left(\frac{D_{\rm nc}}{1.35}\right)^{2.12} \, .
\end{eqnarray}

Finally, the waiting-time distributions obtained in these nonconservative models
show the same overall qualitative behavior as those characterizing
the conservative models (see Figures~\ref{fig:wtdcons} and \ref{fig:Ewtd}).

\subsection{On the Breaking of Finite-Size Scaling\label{ssec:fss}}

In statistical physics parlance, the appearance of several
statistically distinct
populations of avalanches (as occurs here) 
%  at high driving rates and high degree
% of conservation in the redistribution rules)
represents a break
of {\it finite-size scaling}. Such breaks are known to occur in certain
classes of SOC models. For instance, the \citet{Olami:1992do} model for
earthquake, which is also a globally and deterministically driven
model, shows a clear break in finite-size scaling when its
nonconservation parameter (equivalent to $1-D$ herein) becomes too small
\citep[see][]{Grassberger:1994fv}.
This can be traced to the effect of
boundaries on internal avalanche dynamics, with lattice nodes
located near the boundaries becoming favored triggering sites
for large avalanches \citep[see][]{Lise:2001du}. We have searched
for spatial dependencies in the triggering sites of large avalanches
in a few simulation runs, by constructing 2D spatial frequency distributions
functions for onset locus of large and small avalanches. No obvious
dependencies on the distance to boundaries could be detected.

\begin{figure}
  \centering
  \includegraphics[width=\linewidth]{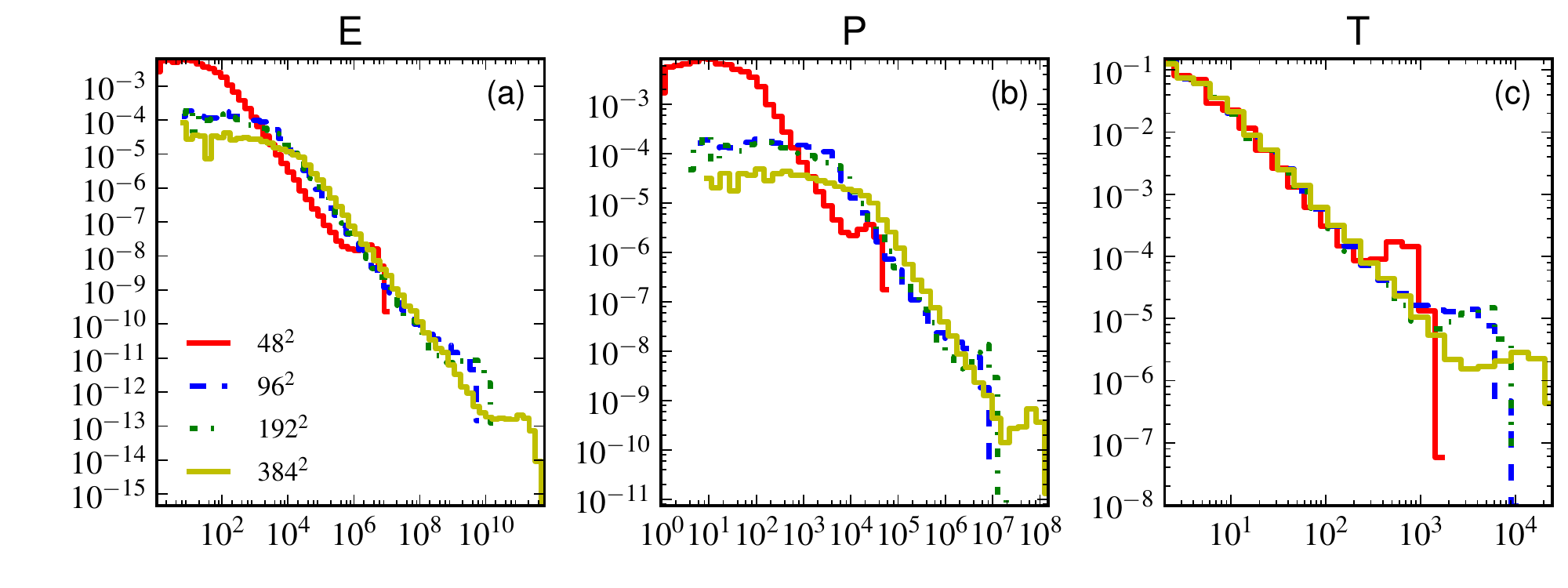}
 \caption{PDFs of E, T, and P for model NC0 [$D_{nc}=0.8$, $\varepsilon
   = 10^{-5}$] for four different lattice sizes.}
  \label{fig:resolution}
\end{figure}

We explored the effect of resolution on the break of finite-size scaling at both
ends of our PDF (Figure \ref{fig:resolution}). We observe that the
population of
large and quasi periodic avalanches always spans the last
decade of energies (both peak and total) for the four sizes of lattice
we considered. It confirms that this population consists of
avalanches spanning the whole lattice and does not result from a
particular scale our model could have introduced. 

The small avalanches population is on the contrary sensitive to the
lattice resolution. The avalanche duration PDF (c) shows no hint of
resolution effect: the small avalanches population results from a
shift of the typical energy an avalanche of small
duration is able to release. We reiterate that the
plateau of small avalanches originates from the fact that small
avalanches have the ability to release energies in a significantly larger range
(spanning higher energies) than the corresponding avalanches of 
the LH model. Calculating the energy [$\Delta e_{NC0}$] released by a one-node and
one-iteration avalanche in model NC0 (Equation (\ref{eq:erel1}), we obtain:
\begin{equation}
  \label{eq:energy_released_NC0}
  \frac{\Delta e_{NC0}}{e_{0}} = \frac{\Delta e_{i,j}}{e_{0}} +
  \frac{1-r_{0}}{2Z_{c}}\sum \left( A_{i\pm 1,j\pm 1}\right) +
  \frac{1-r_{0}^{2}}{5}\, ,
\end{equation}
where we recall that $r_{0}\in [D,1]$. If $r_0=1$, we obtain the
classical LH energy release (Equation (\ref{eq:erel1}). If $r_{0}\approx D$,
the additional energy release is dominated by the second term
-- involving the sum of the nodal variable over the neighbouring
node -- which is of the order of the nodal variable. The worst case
occurs in the middle of the lattice where the 
nodal variable is maximal. A larger lattice leads to
larger values of the nodal variable, which explains the wider small
avalanches populations observed in Figure \ref{fig:resolution}a -- b.
For example, the nodal variable is of the order of $3\times 10^{4}$ for a
lattice of $384^{2}$, which corresponds indeed to the energy limit
distinguishing the small avalanches from the scale-free mid-size ones.
Finally, the range of power-law mid-size avalanches
also increases with resolution, as expected from classical results of
the LH model.

Unlike in
the seismic-fault model (which should be thought of, in some global sense,
as periodic), here coronal loops do have a finite spatial extent
in cross-section, and so the various ``tricks'' that have been
developed to restore finite-size scaling to the Olami \etal~model
\citep[\textit{e.g.}][]{Manna:2006hb} cannot legitimately be used
in the present context.

The model results of \citet{Grassberger:1994fv} suggested that
the loss of finite-size scaling takes place quite suddenly,
beyond a certain level of non-conservation. The results reported
here indicate a more gradual transition,
with the population of large, quasiperiodic avalanches gradually
merging with the mid-size, scale-free avalanches as $D$ decreases
(Figure~\ref{fig:pdfncons}).
% \todo{but this clearly requires further quantitative investigation,
% as the strategy used to distinguish the ``large'' avalanches from
% the high end of the scall-free distribution of ``small'' avalanches
% is not entirely satisfactory.}

\section{Discussion and Conclusion
\label{sec:conclusion}}

In this article, we have described and documented the behavior of a novel
and hitherto unexplored class of lattice-based avalanche models applicable to
solar flares. These models, based on a spatially global
and fully deterministic driving mechanism,
are amenable to physical interpretation in terms of slow twisting of 
a small coronal loop. Stochasticity is introduced in the models at the
level of the redistribution rules and
stability threshold governing avalanching dynamics.

Over significant portions of model space, the simulations produce three
distinct populations of avalanches. The first is composed of the smallest
avalanches with a plateau-like distribution. Then,
a classical scale-free distribution of avalanches 
takes place in the mid-size range of energies. Finally,
quasi-periodic avalanches with a more-or-less
well-defined mean size make up the third
population. The two latter populations are observed to merge in the
context of avalanche models incorporating non-conservative redistribution
rules.%  Even then, scale-free distribution are found to materialize only
% for slow enough driving, and small yet significant levels of dissipation in
% the nodal variable.
 
We have argued that the large, quasiperiodic avalanches are the reflection
of an energy loading/unloading cycle driven by our adopted global deterministic
driving mechanism. Even though an avalanche is always triggered at a single lattice
node, the effect of the global driver is that energy has been accumulating
at {\it all} internal lattice nodes prior to avalanche onset. Once the avalanche begins,
this energy must be either evacuated at the open boundaries, or dissipated locally
within the lattice (in non-conservative models). In conservative models, only
the former is possible, which leads inevitably to the buildup of very large
avalanches that must connect to the boundaries -- and thus end up spanning
the whole system -- and remain active until the excess energy has been drained
away from the lattice -- thus producing long duration avalanches. In that respect,
it is therefore not surprising that a distinct population of large avalanches
(and, consequently, break of finite-size scaling) is most
readily observed in models based on conservative redistribution
rules.

We have been careful thus far to refrain from declaring our models to be
in a true self-organized critical state, using instead the
term ``SOC-like'' to characterize their avalanching behavior.
Formal demonstration of SOC typically requires the computation
of critical exponents, and the demonstration that the latter
satisfy
mutual numerical relations that place them in a specific universality
class. % (see, \textit{e.g.} Uritsky \etal~2001, and references therein).
Although interesting in and of itself, the identification of
the universality class to which the present model belongs
is of limited interest from
the point of view of flare modelling. We stress again that
the models discussed in this article do satisfy all
{\it sine qua non} physical conditions believed to be
required for SOC \citep{Jensen:1998ww}: a slowly driven open system
subjected to a self-limiting, local, finite-threshold instability.
Moreover,
in nonconservative models characterized by no quasi-periodic large
scale events,
the lack of dependence of the power-law indices of avalanches size
and duration on the driving rate and lattice size
certainly suggests that these models all belong to the same
universality class.

Although non-conservative redistribution rules have received, to the
best of our knowledge, little attention in the solar flare context,
true SOC avalanching behavior is known
to materialize in non-conservative system. The most studied such
SOC system is
the cellular automaton formulation of the stick--slip model for earthquake,
introduced by
\citet{Olami:1992do}. In that model, non-conservation is related to the
fact that a moving block exerts a force not only on its neighbour blocks
via the connecting springs, but also on the two plates bounding
(and driving) the system from above and below; hence; when an individual
block moves, a fraction
of the potential energy stored in the springs is dissipated as heat
via friction against the bounding plates,
rather than all of it ending up in other springs by the end of
the avalanche.
Something similar also takes place
in the non-conservative substorm models of \citet{Liu:2006cp} and
\citet{VallieresNollet:2010ed}. There, 
a fraction of the energy released by an unstable magnetic flux tube
within the central (equatorial) plasma sheet is lost via MHD waves
and/or charged particles travelling away along magnetic field lines
back towards Earth,
eventually leading to auroral excitation in polar regions.
A particularly interesting feature of the latter model is its
ability to simultaneously produce ``internal''
avalanches with a scale-free
size distribution, and quasi-periodic boundary discharge avalanches 
with a well-defined mean-size. This squares well with the
statistically distinct properties of auroral emission on the one
hand, and ring current injection events on the other, both being
in principle related manifestation of energy release events in the central
plasma sheet. In the model of \citet{Liu:2006cp}, this effect materialized
in part because the boundaries were set up so as to respond dynamically
to incoming avalanches; the results presented herein indicate
that dual populations of energy release events, each with distinct
statistical characteristics, would still materialize in such
globally driven systems, even with more conventional ``open'' boundary
conditions, provided conservation is high enough. \changed{Finally,
  solar flares also exhibit non-conservative
  properties. Flares are thought to 
  originate from reconnection processes in coronal loops that feed on
  the stored magnetic energy. Along with the magnetic reorganization
  associated with a flare, some energy is systematically lost from the
  system either through radiation (direct X-ray emissions, or
  hard X-ray emissions triggered by reconnection-accelerated energetic
  particles), accelerated particles escape or even conduction along the magnetic 
  field lines. We modeled those effects with a simple non-conservation
  parameter in the avalanching process. Albeit with a quite different
  cellular automaton model, our results are consistent with the
  previous finding of \citet{Hamon:2002hs}: a quasi-cyclic population
  of large avalanches is found in both cases when the model is 
  close to be purely conservative.}

Notwithstanding the presence or absence of a population
of small and large avalanches, in the realm of mid-size avalanches all
models are characterized by a power-law distribution of
avalanche energy release spanning many orders of magnitude.
In general, the corresponding
power-law indices are somewhat
smaller than for the LH model, implying flatter PDFs, so that all models have a 
power-law index smaller than the critical value $\alpha_E=2$ above
which the smaller flares (\textit{i.e.} avalanches) would become the dominant
contributors to coronal heating, as in Parker's nanoflare hypothesis.

An obvious extension of the present model is to introduce the third
spatial dimension; as argued in Section \ref{ssec:phys}, the 2D lattices
considered herein can be best regarded as a cross section
at some fixed position along a coronal loop. Evidently, allowing avalanches
to develop along the loop's length would yield a more realistic
model. In particular, it would become possible to map the
3D lattice to a bent coronal loop, and therefore compute
fractal dimensions of flares/avalanches \citep[see, \textit{e.g.} ][]{Morales:2009dc} in the same manner
as carried out from observations \citep[see,
e.g.][]{Aschwanden:2013kb}. This would open a new comparative bridge
between modelling and observation.

The broad exploration of parameter space made possible
with the computationally much less demanding 2D lattices has
allowed us to pinpoint the most important parameter regime, which is found
for highly dissipative redistribution.
Then, the way in which stochasticity is introduced in the model
(stability threshold, extraction, redistribution rules)
plays a comparatively lesser role in determining the size
distribution of avalanches. In other words, the models are
robust with respect to the details of stochastic effects.

% The organization and storyline of \S\ref{sec:results} may have
% given the impression that
% a second population of large, quasiperiodic avalanches is a 
% nuisance that must be eliminated in order to reproduce observed
% flare size distributions.
% While this may be the case as far as global statistics
% are concerned, the possibility remains that this behavior
% has bearing on recurrent flaring in individual active regions
% or coronal loops. There is observational evidence that a flaring
% active region goes through various stages of flaring growth
% and decay \citep[see][]{LopezFuentes:2007bx}.
% Could this reflect systematic temporal variations
% in driving rates, leading to qualitatively distinct flaring
% behavior over time? This is a question worth further investigation, on both
% the observational and modelling fronts.

The present model finally provides an interesting basis for a
practical use of avalanches models in the context of solar flares
\citep{Strugarek:2014we}. The deterministic driver minimizes the level of
stochasticity embedded in the model, which results in very good predictive
capabilities for the larger events. Interestingly, our model could provide an
efficient and cheap alternative method for the prediction of large solar flares in
the context of space weather.

\begin{acks}
The authors acknowledge stimulating discussions during
the ISSI Workshops on Turbulence and Self-Organized Criticality
(2012 -- 2013) held in Bern
(Switzerland); and during the ``Festival de th\'eorie'' (2013) held in
Aix-en-Provence (France). We acknowledge support from
Canada's Natural Sciences and Engineering Research Council.
\end{acks}

\bibliographystyle{spr-mp-sola}
%\bibliography{../mybib}

\end{article}

\end{document}